\documentclass[fleqn,usenatbib]{mnras}
\usepackage{newtxtext,newtxmath}
\usepackage[T1]{fontenc}
\usepackage{color}

\definecolor{olive}{RGB}{48,188,0}

\DeclareRobustCommand{\VAN}[3]{#2}
\let\VANthebibliography\thebibliography
\def\thebibliography{\DeclareRobustCommand{\VAN}[3]{##3}\VANthebibliography}

\usepackage{graphicx}	
\usepackage{amsmath}	
\usepackage{color}

\newenvironment{tightcenter}{%
	\setlength\topsep{0pt}
	\setlength\parskip{0pt}
	\begin{center}
}{%
 	\end{center}
}

\newcommand\chandra{Chandra}
\newcommand\ciao{CIAO}

\newcommand\caldb{CALDB}

\title[X-ray Observations of Cygnus A-2]{Late-Time X-ray Observations of the Transient Source Cygnus\,A-2}

\author[B. Snios et al.]{
Bradford Snios$^{1}$\thanks{E-mail: bsnios@cfa.harvard.edu},
Martijn De Vries$^{2}$,
Paul E. J. Nulsen$^{1,3}$,
Ralph P. Kraft$^{1}$,
\newauthor Aneta Siemiginowska$^{1}$
and Michael W. Wise$^{4,5}$
\\
$^{1}$Center for Astrophysics $|$ Harvard \& Smithsonian, 60 Garden Street, Cambridge, MA 02138, USA\\
$^{2}$Department of Physics/KIPAC, Stanford University, Stanford, CA 94305, USA\\
$^{3}$ICRAR, University of Western Australia, 35 Stirling Hwy, Crawley, WA 6009, Australia\\
$^{4}$Astronomical Institute ``Anton Pannekoek," University of Amsterdam, Postbus 94249, 1090 GE Amsterdam, The Netherlands\\
$^{5}$SRON Netherlands Institute for Space Research, Sorbonnelaan 2, 3584CA Utrecht, The Netherlands
}

\date{Accepted 2022 February 14. Received 2022 February 11; in original form 2021 November 05}

\pubyear{2022}

\begin{document}
\label{firstpage}
\pagerange{\pageref{firstpage}--\pageref{lastpage}}
\maketitle

\begin{abstract}
We examine \chandra\ observations of the powerful Fanaroff-Riley class II (FR\,II) radio galaxy Cygnus\,A for an X-ray counterpart to the radio transient Cygnus\,A-2 that was first detected in 2011. Observations are performed using the High-Resolution Camera (HRC) instrument in order to spatially resolve Cygnus\,A-2 and the central Active Galactic Nucleus (AGN) at a separation of $0\farcs42$. Simulated images are generated of the emission region, and radial profiles for the region of interest are extracted. A comparison between the simulations and observations reveals no X-ray detection of Cygnus\,A-2 to a 0.5--7.0\,keV flux upper limit of $1.04 \times 10^{-12}\rm\,erg\,cm^{-2}\,s^{-1}$, or a rest-frame 2--10\,keV luminosity of $8.6\times 10^{42}\rm\,erg\,s^{-1}$. We estimate the black hole mass of Cygnus\,A-2 based on our X-ray flux limit and find it to be consistent with a flaring black hole rather than a steadily accreting source. The HRC observations are additionally compared with archival ACIS data from 2016--2017, and both the overall morphology and the flux limits of the AGN complex agree between the two datasets. This consistency is despite the pile-up effect in ACIS which was previously considered to bias the observed morphology of the AGN. The agreement between the datasets demonstrates the viability of utilizing the archival \chandra\ data of Cygnus\,A to analyze its AGN at an unprecedented level of precision.
\end{abstract}




\section{Introduction}
\label{sect:intro}

Amongst Fanaroff--Riley class II radio galaxies \citep[FR\,II;][]{Fanaroff1974}, Cygnus\,A is the archetype of powerful radio galaxies \citep{Carilli1996}. At a redshift of $z = 0.0561$ \citep{Owen1997} and  with an estimated jet power approaching $10^{46}\rm\ erg\ s^{-1}$ \citep[e.g.,][]{Godfrey2013}, it is by far the nearest truly powerful radio galaxy in the universe. Cygnus\,A is hosted by the central galaxy of a rich, cool-core galaxy cluster \citep{Owen1997}, and X-ray observations provide a valuable probe of the energy flows through the jets from its Active Galactic Nucleus (AGN), the interaction of the jets with the surrounding medium, and the overall system's impact on its environment  \citep{Carilli1988, Carilli1994, Harris1994, Smith2002, Rafferty2006, deVries2018, Duffy2018, Snios2018b}. X-ray observations of Cygnus\,A also may be used to investigate the physical structure of a powerful radio galaxy and discuss its evolution over time. Beyond the factors that make Cygnus\,A interesting on its own, analysis of this system is also beneficial to the study of FR\,II systems in general.

In 2015, a new radio source was observed in Cygnus\,A with the Karl G. Jansky Very Large Array \citep[VLA;][]{Perley2017}. The transient source, named Cygnus\,A-2, is offset from the Cygnus\,A nucleus by a projected distance of 0\farcs42 to the southwest. Follow-up observations in 2016 with the Very Long Baseline Array (VLBA) confirmed the presence of the source and showed that it had not varied significantly in brightness within that year. Cygnus\,A-2 was not detected in archival VLA observations in between 1989 and 1997. A follow-up study by \cite{Tingay2020} reviewed VLBA archival data of the system from 2002--2013, wherein Cygnus\,A-2 was detected in a 15.4\,GHz observation from 2011. These results establish that the flux density of Cygnus\,A-2 increased by at least a factor of five in a span of 4--6 years.

One of the most distinct features of Cygnus\,A-2 is its radio spectral luminosity of $L_{\nu} \approx 3 \times 10^{29}$\,erg\,s$^{-1}$\,Hz$^{-1}$, or $\nu F_{\nu} \approx 6 \times 10^{39}$\,erg\,s$^{-1}$, making it a very luminous radio transient \citep{Pietka2014}. Based on its radio luminosity, three possible origin mechanisms were considered by \cite{Perley2017}. The first possibility is that Cygnus\,A-2 is a supernova, although the high luminosity excludes the majority of common supernovae subclasses. A relativistic, gamma-ray burst supernova could achieve the required radio luminosity, but these events are very rare \citep{Perez-Torres2015}. Additionally, Cygnus\,A-2 coincides with an infrared point source previously observed in 2003 by the Keck Observatory \citep{Canalizo2003}, the intensity of which is difficult to explain with current supernova models.

The second possibility is that Cygnus\,A-2 is a secondary AGN, in orbit around the primary AGN of Cygnus\,A. The coincidence with the infrared point source can then be explained as light from the AGN itself, or as light from a tidally stripped galaxy remnant, of which the secondary AGN sits at the center \citep{Canalizo2003}. The appearance of the AGN over the 18-year timespan could have been caused by a steady rise in luminosity, as Cygnus\,A-2 came into the interaction range of accreting material. However, the sensitivity upper limits in the archival VLBA observations imply that it has increased in brightness by at least a factor of five \citep{Tingay2020}. Such a significant change in brightness is very large for a non-blazar AGN. Previous studies by \cite{Hodge2013} and \cite{Wolowska2021} on blindly selected radio sources found that less than 1\% of their targets varied by factors greater than three over a decade timescale. Therefore, the probability is low that such a scenario caused the significant brightness change observed from Cygnus\,A-2. 

Alternatively, if Cygnus\,A-2 is a secondary black hole, it could have undergone sudden large-scale accretion through a Tidal Disruption Event (TDE). The brightest TDEs, such as Swift J1644$+$57 \citep{Burrows2011, Zauderer2013}, have reached radio luminosities up to $10^{42}$ erg s$^{-1}$, significantly more than the $ 6 \times 10^{39}$ erg s$^{-1}$ observed in Cygnus\,A-2. TDEs are rare, happening at estimated rates of $10^{-4}$--$10^{-5}$ yr$^{-1}$ galaxy$^{-1}$, although theoretical modeling suggests that disruption rates can be significantly enhanced in systems with two black holes in close orbit \citep{Liu2013}. Regardless of which explanation, if any, is shown valid for the origin of Cygnus\,A-2, it is clear that the uniqueness of this source merits further study.

\begin{table}
\caption{Overview of \chandra\ Observations}
\begin{tightcenter}
\begin{tabular}{c c c c c c }
\hline \hline
Date & ObsID & Instrument & Roll Angle & $T_{\rm exp}$ & $C_{\rm obs}${}$^{a}$\\ 
(UTC) &   &  & (deg) & (ks) & \\
\hline
2016-06-13 & 18871 & ACIS & 146 & 21.6 & 2853 \\
2016-06-18 & 17133 & ACIS & 146 & 30.2 & 3930 \\
2016-06-26 & 17510 & ACIS & 156 & 37.1 & 5011 \\
2016-07-10 & 17509 & ACIS & 170 & 51.4 & 7053 \\
2016-08-15 & 17513 & ACIS & 209 & 49.4 & 6818 \\
2017-01-20 & 17135 & ACIS & ~~~8 & 19.8 & 2623 \\
2017-01-26 & 17136 & ACIS & ~~~8 & 22.2 & 2950 \\
2017-01-28 & 19996 & ACIS & ~~~8 & 28.1 & 3722 \\
2017-02-12 & 19989 & ACIS & ~~30 & 41.5 & 5446 \\
2017-05-10 & 17511 & ACIS & 105 & 15.9 & 2034 \\
2017-05-13 & 20077 & ACIS & 105 & 27.7 & 3500 \\
2017-05-20 & 17134 & ACIS & 126 & 28.5 & 3184 \\
2017-05-21 & 20079 & ACIS & 126 & 23.8 & 2976 \\
2021-05-04 & 22536 & HRC & 100 & 19.9 & 1594 \\
2021-05-05 & 25020 & HRC & 110 & 19.9 & 1636 \\
2021-05-07 & 25021 & HRC & 112 & 19.9 & 1593 \\
\hline
\end{tabular}
\label{table:obs}
\end{tightcenter}
${}^{a}$\,Total observed counts from a 1\farcs5 radius circle centered on the Cygnus\,A AGN. Counts for ACIS images are over the 0.5--7.0\,keV band, while HRC counts are for the 0.08--10.0\,keV band.
\end{table}

Recently, \cite{deVries2019} analyzed archival \chandra{} observations of Cygnus\,A taken between 2015--2017 with the Advanced CCD Imaging Spectrometer (ACIS) to search for evidence of an X-ray counterpart to Cygnus\,A-2. At a separation of $0\farcs42$ between the AGN and Cygnus\,A-2, the transient source cannot be spatially resolved with ACIS. The data were instead compared with simulated \chandra\ observations of the system generated using the {\tt MARX} software package. Moderate pile-up was inferred in the ACIS observations of the AGN, which required additional spectroscopic and morphological modeling for the simulations. Results from the X-ray data provide a non-detection of the transient counterpart, placing an upper limit to the 2--10\,keV X-ray luminosity of Cygnus\,A-2 of $1 \times 10^{43} \rm\ erg\ s^{-1}$. The X-ray flux upper limit for Cygnus\,A-2 disfavours the interpretation of Cygnus A-2 as a steadily accreting black hole, leaving a TDE as the most likely cause.

Although the result from \cite{deVries2019} is a valuable constraint, the accuracy of the flux limit could be improved if spatial resolution is increased and pile-up issues are removed. Among the available X-ray instruments, \chandra's High-Resolution Camera (HRC) is ideal for such a study.  HRC provides the densest sampling of the High Resolution Mirror Assembly (HRMA) point-spread function (PSF) available on Chandra, with a pixel size of 0\farcs132 and a nominal point spread with full width at half maximum of $0\farcs4$. The microchannel plate detector on HRC is also unaffected by pileup, in contrast to ACIS. It is therefore the focus of this work to examine \chandra\ HRC observations of the Cygnus\,A system to more accurately quantify the X-ray emission properties of the TDE candidate Cygnus\,A-2. 

The remainder of the paper is structured as follows. Section~\ref{sect:obs} describes the X-ray observations of Cygnus\,A and data processing. Section~\ref{sect:radial} describes the radial profile analysis of the AGN core, with an emphasis on detecting the nearby source Cygnus\,A-2. Section~\ref{sect:discuss} details the measured properties of Cygnus\,A-2 as determined from the HRC data, and it is additionally discussed in context of the previous X-ray results. Section~\ref{sect:mass} estimates a black hole mass for Cygnus\,A-2, which is used to infer emission properties of the source. Section~\ref{sect:hrc} explores implications of studying the Cygnus\,A core with the \chandra\ HRC instrument, and our concluding remarks are provided in Section~\ref{sect:conclude}. 

For this paper, we adopted the cosmological parameters $H_{0} = 70\rm\,km\,s^{-1}\,Mpc^{-1}$, $\Omega_{\Lambda} =0.7$, and $\Omega_{M} = 0.3$ \citep{Hinshaw2013}, giving an angular scale for Cygnus\,A of $1.103\rm\ kpc\ arcsec^{-1}$ and an angular diameter distance of 227\,Mpc at the redshift $z = 0.0561$. 

\section{Data Analysis}
\label{sect:obs}

\subsection{HRC Observations}

\begin{figure}
\begin{tightcenter}
\includegraphics[width=0.995\linewidth]{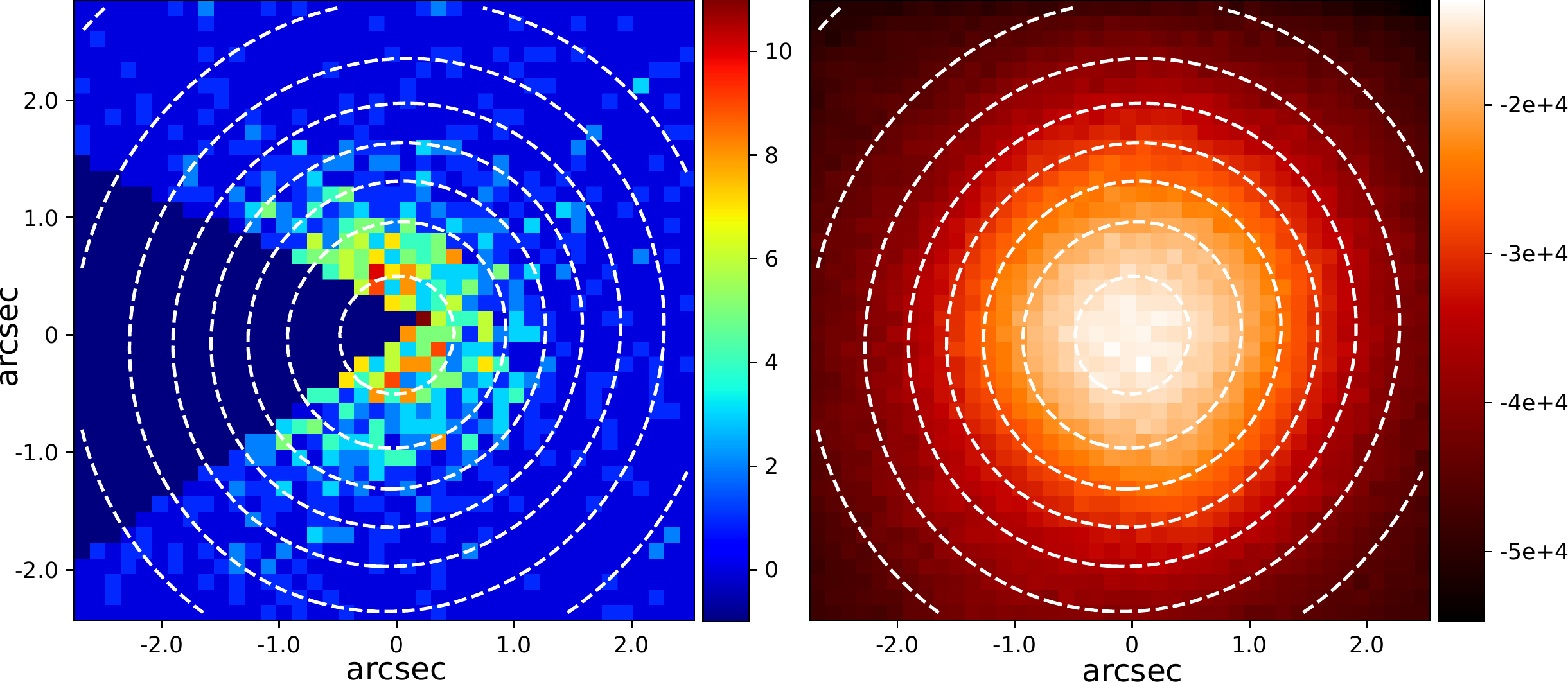}
\caption{An example of the `Figure of Merit' method for ObsID\,22536. {\it Left panel:} the \chandra\ observation of the AGN core, binned at native HRC pixel resolution (0\farcs132 per pixel), and a colorbar indicating the number of counts in each pixel. A 60$^{\circ}$ wedge around the PSF artifact is masked to avoid biasing the resulting best-fit. {\it Right panel:} the constructed Figure of Merit, showing the normalized likelihood, $P( {\rm data} | {\rm model})$, as a function of pixel shift. The dashed contours trace the fitted 2-D Gaussian, which are also overplotted on the observation image. Tick labels indicate the distance in arcsec away from the fitted center.}
\label{fig:FoM}
\end{tightcenter}
\end{figure}

Observations of Cygnus\,A were performed with the \chandra\ \mbox{HRC-I} instrument with the AGN centered at the aimpoint to minimize the size of the PSF. A complete list of the observations is shown in Table~\ref{table:obs}. Each observation was reprocessed using \ciao\,4.13 and \caldb\,4.9.5 \citep{Fruscione2006}. Roll angle for each observation was chosen such that the TDE did not coincide with the known \chandra\ PSF artifact\footnote{See the PSF artifact caveat in the \ciao{} User Guide \\ \url{https://cxc.harvard.edu/ciao/caveats/psf_artifact.html}} that creates asymmetric X-ray features on a subarcsecond scale, and we verified that the azimuth of the artifact extension is at least $60^\circ$ SE of the azimuth of Cygnus\,A-2 relative to the nucleus.

Radio observations of Cygnus\,A measured a separation of $0\farcs42$ between the AGN and Cygnus\,A-2, which is comparable to the $0\farcs4$ nominal PSF full width at half maximum (FWHM) of the HRC instrument. Thus, extra care must be taken in aligning the observations in order to analyse the PSF width of the co-added observations. We therefore applied the `Figure of Merit' method. This method is described in more detail in \cite{vanEtten2012, deVries2021}, and briefly summarized here. We created simulations in {\tt MARX} of the AGN, placing the source at the astrometric position of the radio nucleus, $\alpha_{\rm nuc}=$ 19:59:28.35648, $\delta_{\rm nuc}= $ +40:44:02.0963 \citep{Gordon2016}. The simulations were done following \cite{deVries2019}, by combining simulations of the AGN and the 0.5--2.2\,keV scattering clouds that directly surround the AGN. We then cut out images of the data and the simulation at 0\farcs066 pixel size (0.5$\times$ native HRC resolution).  The Figure of Merit is constructed for each individual exposure by shifting the simulated data in the $x$ and $y$ direction, and calculating the Poisson likelihood, $P({\rm data} | {\rm model})$, at each pixel shift. The Figure of Merit thus maps the pixel shifts at which the simulation provides the best match to the data. Finally, we determine the centroid by fitting a 2-D Gaussian to the Figure of Merit, and subsequently apply the appropriate pixel shifts to the data such that they match the astrometric position of the radio nucleus. 

\begin{figure}
\begin{tightcenter}
\includegraphics[width=0.99\linewidth]{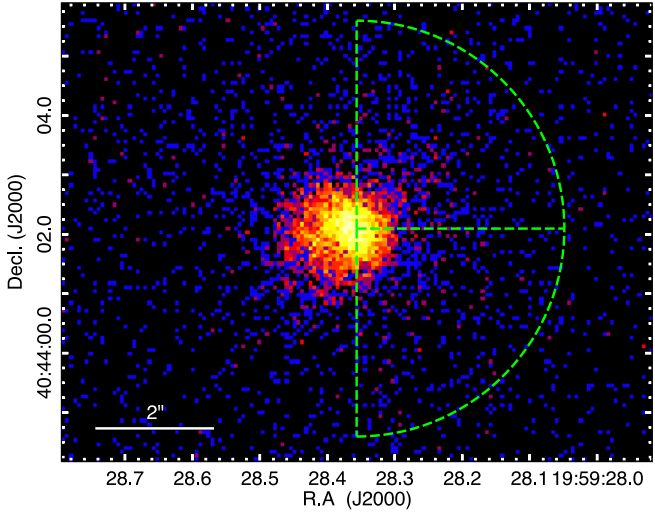}
\caption{A co-added, HRC-I \chandra\ image of the Cygnus-A core. The image includes all observations from Table~\ref{table:obs} after having performed astrometric alignments, and the pixel size is 0\farcs066. The radial profile regions (green, dashed) used in Section~\ref{sect:radial} are shown. Extended X-ray emission observed to the southeast of the core is due to the known \chandra\ PSF artifact and is therefore nonphysical.}
\label{fig:cyga}
\end{tightcenter}
\end{figure}

While constructing the Figure of Merit, we mask pixels in a $60^{\circ}$ wedge around the PSF artifact to avoid any potential bias in our determined centroid. An example figure of merit and fitted position is shown in Figure~\ref{fig:FoM}. The reprojected dataset was co-added with the {\tt merge\_obs} routine, and the merged 60\,ksec image of the three HRC observations is shown in Figure~\ref{fig:cyga}.

\subsection{ACIS Observations}

Since the flux of Cygnus\,A-2 is known to vary in the radio \citep{Perley2017}, we investigated for evidence of similar flux changes in X-rays. We obtained a set of \chandra\ ACIS observations taken between June 2016 and May 2017 to compare against the HRC data. The selected ACIS observations are described in more detail in \cite{deVries2019}, wherein no evidence of Cygnus\,A-2 was detected. For consistency, we reprocessed the observations with \ciao\,4.13 and \caldb\,4.9.5 and performed the astrometric alignment using the Figure of Merit method in the same manner as for the HRC data.

As in the HRC observations, the \chandra\ PSF artifact can be observed in the ACIS data, where the artifact is dependent on the roll angle of the satellite. Since the ACIS observations were taken over a large range of roll angles, the PSF artifact coincides with the location of Cygnus A-2 in some of the archival ACIS data. We additionally require a `clean' radial profile of the Cygnus\,A core with which we may compare against the Cygnus\,A-2 region. Thus, we created two separate, co-added ACIS images: one image of all observations in which the artifact is not in the SW quadrant (the region where Cygnus\,A-2 resides), and one image of all observations in which the artifact is not in the NW quadrant (a `clean' region for reference). From available ACIS data, this criterion equates to a roll angle greater than $60^\circ$ for the SW quadrant (total 287\,ks), and a roll angle less than $150^\circ$ for the NW quadrant (total 261\,ks). Consistent with our HRC analysis, a $60^{\circ}$ wedge around the PSF artifact was masked when determining the centroid, and the reprojected data sets were co-added with the {\tt merge\_obs} routine. A list of the utilized ACIS observations is provided in Table~\ref{table:obs}. 

\section{Radial profiles of Cygnus\,A AGN}
\label{sect:radial}

\begin{figure}
\centering
\includegraphics[width=0.99\linewidth]{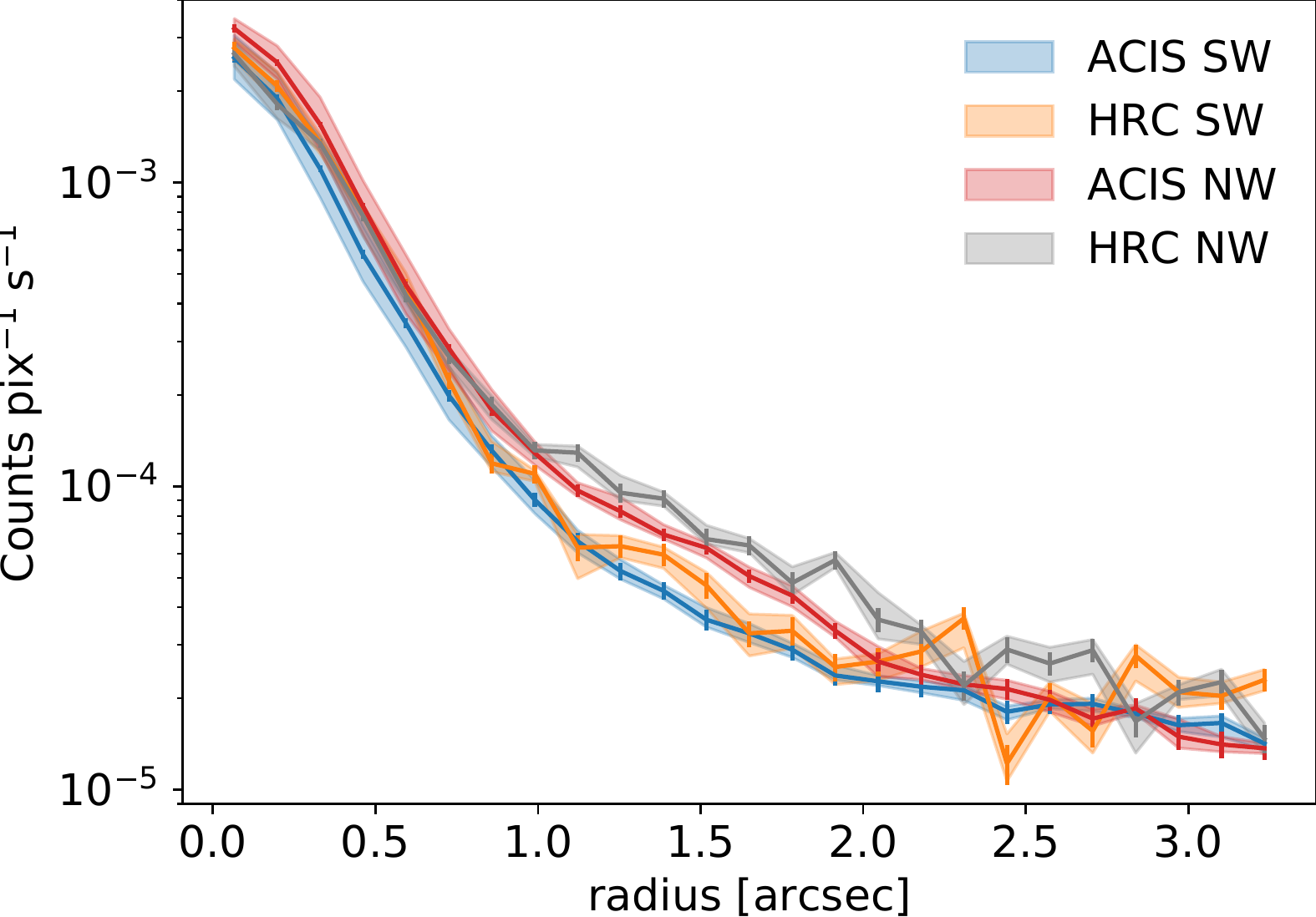} \\
\caption{Radial profiles of ACIS and HRC NW and SW quadrants. HRC counts were scaled to match the ACIS counts in the same quadrant at 3\arcsec, in order to correct for the different effective areas. The error envelope was done through a Monte-Carlo procedure, and reflects the uncertainty in the centroid location, as determined from the Figure of Merit. The error bars on the median lines reflect the Poisson error.}
\label{fig:rprofiles}
\end{figure}

Using the HRC and ACIS co-added images from Section~\ref{sect:obs}, we created radial profiles of the AGN region in order to measure whether 1) the SW quadrant (with the transient) is more extended, and 2) whether a profile change can be observed between the HRC and ACIS observations. Radial profiles were extracted with the {\tt dmextract} routine in \ciao, where a 0.25 pixel (0\farcs123) binning was used for the ACIS data in order to match the pixel size of HRC. Radial profiles were extracted for the SW and NW quadrants of the images, and the profiles are shown in Figure \ref{fig:rprofiles}.

Due to different energy responses of the ACIS and HRC instruments, we opted to not exposure-correct the final radial profiles to avoid possible biasing. Instead, the X-ray surface brightness over the 3--5\arcsec\ radial distance from the AGN core was set equal between the different profiles, allowing us to compare relative changes in the morphology between the different quadrants. An error envelope for each profile was estimated through a Monte-Carlo procedure wherein we varied the location of the centroid using the uncertainties obtained from the Figure of Merit procedure described in Section~\ref{sect:obs}. Overall, a similar uncertainty was found for each quadrant. 

Examining the ACIS and HRC radial profiles shows that the NW and SW quadrants are consistent between the two instruments, indicating that the radial profile of the Cygnus\,A AGN is constant between the 2016--2017 and 2021 observations. Comparing the different quadrants also reveals that the SW radial profile at $0\farcs42$, where Cygnus\,A-2 is located, agrees with the `reference' NW profile within 1$\sigma$, suggesting that no X-ray emission from the possible TDE source was detected in either the ACIS or HRC observations. Additionally, an X-ray flux enhancement is observed in the NW quadrant at a distance of $\sim$\,1\farcs5 from the core that is not present in the SW. Implications of these results are discussed in the remainder of this article.
\section{X-ray Properties of Cygnus\,A-2}
\label{sect:discuss}

\begin{figure}
\centering
\includegraphics[width=0.99\linewidth]{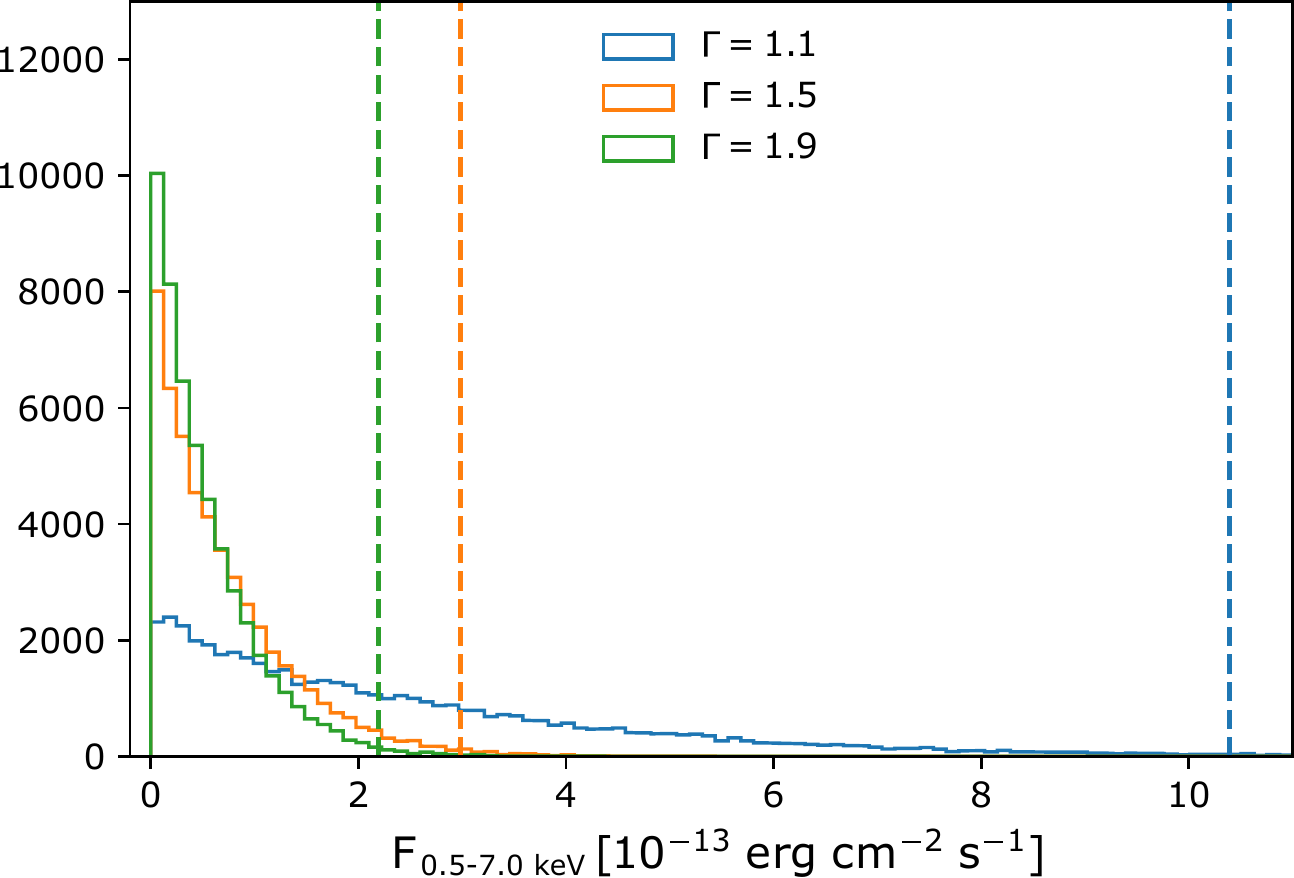} \\
\caption{The posterior distributions obtained from MCMC sampling for the parameter $\alpha$, converted to a 0.5--7.0\,keV flux. Shown are the distributions for input spectra with photon indices $\Gamma$ of 1.1, 1.5, and 1.9. The corresponding vertical dashed lines indicate the 99$^{\rm th}$ percentile of each distribution.}
\label{fig:post}
\end{figure}

Comparing the NW and SW quadrants of the HRC image in Figure~\ref{fig:rprofiles} shows that the number of counts as a function of radius are equal to each other within the Poisson errors. Thus, there is no evidence for an enhancement in the SW quadrant at $0\farcs42$ where the radio transient source Cygnus\,A-2 is located. The result is consistent with the radial profiles from the archival ACIS observation from 2016--2017, from which \cite{deVries2019} previously demonstrated no transient X-ray source. Thus, the transient Cygnus\,A-2 was not detected in X-rays with \chandra\ on either short- or long-term timescales. 

We may place an upper limit on the flux of Cygnus A-2 by estimating the flux required to produce a statistically significant enhancement in the SW radial profile above the Poisson errors in several radial bins. The statistical significance is calculated through the cumulative Poisson probability $P(N_{\rm SW} \leq M_{\rm AGN} + M_{\rm CygA2})$, where $N_{\rm SW}$ represents the number of counts in the SW quadrant, $M_{\rm AGN}$ the model-predicted counts for the AGN, and $M_{\rm CygA2}$ the model-predicted counts for Cygnus A-2.

Because the AGN is time-variable, obtaining an accurate model through simulations is challenging. We therefore used the number of counts from the NW quadrant as a proxy for the model-predicted AGN counts. In order to obtain the model-predicted counts for Cygnus\,A-2, we generated simulated observations with \texttt{MARX} using an absorbed power law model with photon indices $\Gamma$ of 1.1, 1.5, and 1.9. Since the total absorption column density is known to vary across Cygnus\,A \citep[e.g.,][]{deVries2018,Snios2020}, the column density along the line of sight for Cygnus\,A-2 is not well-known. We therefore set the total absorption column density $N_{\rm H}=5\times10^{21}$\,cm$^{-2}$. We note that the column density parameter directly impacts the observed soft X-ray flux, where increases in $N_{\rm H}$ cause decreases in the observed flux at energies $< 1.0$\,keV. Thus, the column density parameter will more significantly impact spectral models where soft X-ray emission dominates, such as those with larger photon indices. Radial profiles were created from the simulations in the same manner as for the HRC data (Section~\ref{sect:radial}), and the simulated profiles were compared to the observations assuming Poisson errors. 

\begin{table}
\caption{Measured Physical Limits of Cygnus\,A-2}
\begin{tightcenter}
\begin{tabular}{c c c c }
\hline
Model$^{a}$ & $F_{\rm 0.5-7.0\,keV}$ & ${L_{\rm 2-10\,keV}}^{b}$ & ${M_{\rm BH}}^{c}$ \\ 
 & ($10^{-13}\rm\,erg\,cm^{-2}\,s^{-1}$) & ($10^{42}\rm erg\,s^{-1}$) & ($10^{8}\,M_{\odot}$) \\
\hline \hline
\vspace{0.5em}
$\Gamma = 1.1$ & $<10.4$ & $<8.6$ & $>1.3_{-1.1}^{+6.3}$ \\ 
\vspace{0.5em}
$\Gamma = 1.5$ & $<3.0$ & $<2.0$ & $>3.0_{-2.4}^{+11.4}$ \\ 
$\Gamma = 1.9$ & $<2.2$ & $<1.1$ & $>4.3_{-3.3}^{+14.5}$ \\ 
\hline
\end{tabular}
\label{table:measurements}
\end{tightcenter}
${}^{a}$\,X-ray flux upper limits are estimated from simulated observations generated in {\tt MARX}. The simulations assume an absorbed power law emission spectrum with photon indices $\Gamma$ of 1.1, 1.5, and 1.9. ${}^{b}$\,Rest-frame 2--10\,keV luminosity. ${}^{c}$\,Black hole mass estimated with Equation~\ref{eq:mass}, the black hole fundamental plane mass predictor function of \cite{Gultekin2019}. 
\end{table}

Because the simulated profiles are themselves Poisson realizations  of an underlying model, we repeated the  simulations $300$ times and took the median in a given radial bin to estimate the number of Cygnus A-2 counts predicted by the model. We then constructed the following likelihood function:
\begin{equation}
    \mathcal{L} = \sum_{i=0}^{J} \log P (N_{\rm i,SW} \leq M_{\rm i,NW} + \alpha M_{\rm i,CygA2}),
\end{equation}
where J is the total number of bins being summed over, and P is the cumulative Poisson probability that the number of counts from the data in a given bin N$_{\rm i,SW}$ is smaller than the model-predicted counts in that bin. The parameter $\alpha$ is a free parameter representing a normalization scaling factor for the counts from Cygnus A-2, given an input absorbed power law spectrum with an absorption and photon index. By measuring the posterior distribution of $\alpha$, we can infer the relative likelihood of different flux levels and thus place an upper limit to the flux of Cygnus\,A-2 at some statistical significance.

The NW profile deviates from the SW profile beyond $\sim$\,1\arcsec\ because of the presence of the electron scattering clouds along that direction (see Section~\ref{sect:hrc} for further discussion). Therefore, we only used the inner six radial bins to calculate $\mathcal{L}$. To obtain the posterior distribution we use the affine-invariant Markov Chain Monte Carlo sampler of \cite{Goodman2010}, which is implemented through the Python package \textit{emcee}. Figure~\ref{fig:post} shows the posterior distributions for $\alpha$, converted to 0.5--7.0\,keV flux, as well as the 99$^{\rm th}$ percentiles, for the three assumed spectra.

The estimated 0.5--7.0\,keV flux and rest-frame 2--10\,keV luminosity limits from the posterior distributions are shown in Table~\ref{table:measurements}. As is to be expected, the distributions depend on the assumption of the input spectrum. We found that the 99$^{\rm th}$ percentile upper limit from the hardest spectrum of $\Gamma=1.1$ corresponded to an observed \mbox{0.5--7.0\,keV} flux of $10.4 \times10^{-13}$\,erg\,cm$^{-2}$ s$^{-1}$, while the softest spectrum sampled of $\Gamma=1.9$ is a factor of $\sim$\,5 lower at $2.2\times10^{-13}$\,erg\,cm$^{-2}$\,s$^{-1}$. Despite the range in values, all flux limits obtained from our analysis were consistent with those found in \cite{deVries2019} for the ACIS data. Furthermore, when using the same emission model as  \cite{deVries2019} of an absorbed power law with a photon index $\Gamma=1.5$, the measured flux upper limit from the HRC observations is lower than their estimate by a factor of 4. Thus, our measurements provide the best limits to date of the late-time X-ray activity profile for Cygnus\,A-2 while also extending the timespan over which the X-ray emission from this potential TDE is constrained.

\section{Mass Estimation of Cygnus\,A-2}
\label{sect:mass}

Using the new X-ray flux upper limits, we can repeat the analysis from Section\,5.2 of \cite{deVries2019}, and estimate the mass of the Cygnus A-2 black hole if it were steadily accreting. To accomplish this task, we used the black hole fundamental plane mass predictor function of \cite{Gultekin2019}:
\begin{equation}
\begin{aligned}
\label{eq:mass}
    \log(M/10^8 M_{\odot})= & (0.55 \pm 0.22) + (1.09 \pm 0.10 )\log(L_R / 10^{38} \rm{erg\,s^{-1}}) \\
    & - (0.59_{-0.15}^{+0.16}) \log(L_x / 10^{40} \rm{erg\,s^{-1}} ), \\
\end{aligned}
\end{equation}
where $L_R$ is the 5\,GHz continuum radio luminosity and $L_X$ is the \mbox{2--10\,keV} X-ray luminosity. We used the X-ray luminosity upper limits from our different emission spectrum models with the detected radio luminosity from \cite{Perley2017}, and the resulting minimum black hole mass estimates are shown in Table~\ref{table:measurements}. We note that our quoted errors do not include the Gaussian intrinsic scatter on the fundamental plane equation of $\rm ln\,\epsilon_\mu = -0.04\substack{+0.14\\-0.13}$ from \cite{Gultekin2019}, so our errors should be regarded as lower estimates.

From our X-ray luminosity limits, we determined that Cygnus\,A-2 must have a minimum black hole mass of $1.3_{-1.1}^{+6.3} \times 10^{8} M_{\odot}$ assuming it is a steadily accreting source. Our limit is consistent with the minimum mass estimate of $4\times 10^{8} M_{\odot}$ from \cite{deVries2019}, despite the use of a different X-ray luminosity limit and a fundamental plane equation in the previous study. For completeness, we repeated our mass calculation using the upper limit from \cite{deVries2019}, and we found a mass limit of $1.2_{-1.0}^{+5.9} \times 10^{8} M_{\odot}$ which agrees with our analysis. The high black hole mass limit together with the low X-ray luminosity limit disfavors the steadily accreting black hole interpretation of Cygnus\,A-2 as the source would be a significant outlier when compared against standard AGN populations \citep[e.g.,][]{Plotkin2012,Gultekin2019}. Our result therefore suggests that Cygnus A-2 is a flaring black hole.

\section{Morphology of Cygnus\,A Core}
\label{sect:hrc}

Our comparison of the ACIS and HRC radial profiles performed in Section~\ref{sect:radial} demonstrates excellent consistency of the Cygnus\,A core between the two instruments after accounting for the \chandra\ PSF artifact. All profiles are in agreement within 3$\sigma$ out to a radius of 3\farcs5, at which point the thermal emission from the surrounding intracluster medium dominates the observed flux. The agreement of the profiles over the 6 years investigated in this study also suggests a relatively constant power output from the Cygnus\,A core, as any variations would likely manifest as changes of the PSF size. This steadiness in the emitted flux of Cygnus\,A was also found in ACIS observations ranging between 2003--2017 \citep{deVries2019}. Thus, the AGN of Cygnus\,A is a remarkably steady X-ray source in recent years.

In addition to the overall agreement of the radial profiles, both ACIS and HRC observations of Cygnus\,A show evidence of an X-ray flux enhancement in the NW quadrant at a distance of $\sim$\,1\farcs5 from the core. This feature is significant at a $>5\sigma$ confidence level in both the HRC and the ACIS data. Dust extinction is known to be present across the NE-SW direction of the AGN due to scattering clouds \citep{Young2002,deVries2019}, which creates comparatively bright soft X-ray emission in the NW and SE quadrants of the AGN at similar radial distances to those measured in our analysis. Since soft X-ray emission is detected with both the HRC and the ACIS detectors, the elevated flux in the NW quadrant is consistent with the relative increase in soft X-ray emission due to the scattering clouds effect. We also verified that our simulated PSF image shows no evidence of enhancement in the NW quadrant at the observed roll angles, further motivating that this emission is physical in origin. Thus, the observed elevated emission in the NW quadrant is consistent with scattering clouds in the Cygnus\,A core that cause asymmetric soft X-ray flux from the AGN.  

The similarity of the radial profiles is also valuable when studying instrument differences between the various \chandra\ detectors. Historically, ACIS is known to suffer from moderate pile-up when observing the Cygnus\,A core \citep[e.g.,][]{Duffy2018, deVries2019}, where the measured pile-up fraction is $\approx$\,$20\%$. The observed pile-up level ultimately limits the scope of morphological and spectroscopic studies that can be performed on the AGN. In particular, pile-up in the ACIS detector is known to create asymmetry in the wings of the PSF due to charge transfer inefficiencies\footnote{See the Overview of Pileup in the CIAO User Guide\\ \url{https://cxc.harvard.edu/ciao/why/pileup\_intro.html}} which introduces potential biases to the observed radial profile of the AGN. However, HRC does not experience pile-up, giving us a `true' look at the core's morphology with these latest observations. 

In comparing the HRC and ACIS radial profiles of the AGN core in Cygnus\,A, we found the two to be in excellent agreement with one another (Figure~\ref{fig:rprofiles}). The consistency between the two observations indicates that the moderate pile-up detected in previous ACIS observations, with an estimated pileup fraction of $\approx$\,20\% \citep{deVries2019}, does not impact the observed morphology of the AGN. Thus, our results demonstrate the viability of utilizing archival ACIS data totalling 2\,Ms to study the Cygnus\,A core morphology at high-precision, after PSF artifact corrections. Such a study would probe the Cygnus\,A core at a currently unprecedented level of precision, though such an analysis is beyond the scope of our current study. 

\section{Conclusion}
\label{sect:conclude}

We analyzed \chandra\ HRC observations, totalling 60\,ksec, of the powerful FR\,II radio galaxy Cygnus\,A for evidence of the transient radio source Cygnus\,A-2 at a separation of 0\farcs42 from the central AGN. Images were co-aligned to high-accuracy, and radial profiles for the region of interest were extracted. Simulated images of the AGN were generated using \chandra\ PSF models, and we compared the simulations and observations for evidence of extended X-ray emission. From our image analysis, we detected no X-ray emission from the transient source Cygnus\,A-2. Using \texttt{marx} simulations, we determined that an observed 0.5--7.0\,keV flux of $1.04 \times 10^{-12}\rm\,erg\,cm^{-2}\,s^{-1}$, or a rest-frame 2--10\,keV luminosity of $8.6\times 10^{42}\rm\,erg\,s^{-1}$, would correspond to a significant (99$^{\rm th}$ percentile) enhancement in the SW quadrant, which we concluded to be the upper flux limit of Cygnus\,A-2.

Archival \chandra\ ACIS observations were analyzed to investigate if the transient radio source Cygnus\,A-2 varied in brightness over time, where all ACIS images were processed similarly to the HRC data. Comparing the HRC results to the non-detection from archival ACIS data taken between 2016--2017, we found the morphology and flux measurements to agree between the two datasets. Under the assumption that Cygnus\,A-2 is a black hole, we used our derived flux limit to estimate its mass. We found that Cygnus\,A-2 would need to be excessively massive for it to be a steadily accreting source ($>$\,$10^8 M_{\odot}$), thereby favoring a flaring black hole interpretation. Altogether, we conclude that Cygnus\,A-2 is a probable flaring source that was not detected in X-rays on either short- or long-term time scales after its initial radio detection in 2011.

Beyond the analysis of Cygnus\,A-2, we explored the morphology of the AGN in Cygnus\,A using both the HRC and ACIS \chandra\ instruments. An emission feature is observed in the NW quadrant at a distance of $\sim$\,1\farcs5 from the AGN at a $>5\sigma$ level, which we attribute to the asymmetric scattering clouds within the Cygnus\,A core. Overall, we found the AGN morphology to be remarkably comparable between the two instruments out to a radial distance of 3\farcs5. This is surprising given that ACIS is known to suffer from moderate pile-up when observing Cygnus\,A, which was assumed significant enough to limit previous morphological studies. However, the agreement of the ACIS images with the HRC data, which is devoid of pile-up, demonstrates that pile-up provides a negligible impact to the observed morphology of the AGN. Thus, future studies may use the complete collection of archival Cygnus\,A observations, totalling 2\,Msec, to examine the AGN morphology at an unprecedented level of precision. 

\section*{Acknowledgements}

We thank the referee for their comments and suggestions. Support for this work was provided by the National Aeronautics and Space Administration through \chandra{} Award Numbers GO8-19093X and GO0-21101X issued by the \chandra\ X-ray Observatory Center, which is operated by the Smithsonian Astrophysical Observatory for and on behalf of the National Aeronautics Space Administration under contract NAS8-03060. P.E.J.N., R.P.K. and A.S were supported in part by NASA contract NAS8-03060.

\section*{Data Availability}

Observation IDs (ObsIDs) for all data used in this investigation are provided in Table~\ref{table:obs}, and the data is available online in the Chandra Data Archive. The processed data utilized in this article will be made available under request to the corresponding author.

\bibliographystyle{mnras}
\bibliography{all_data}

\bsp	
\label{lastpage}
\end{document}